\documentclass[12pt, preprint]{emulateapj}
\usepackage{natbib}
\usepackage{mathrsfs}
\usepackage{url}
\usepackage{longtable}

\newcommand{\kms}{\ifmmode {\rm km~s}^{-1} \else km~s$^{-1}$\fi}
\newcommand{\ergs}{\ifmmode {\rm erg~ s}^{-1} \else erg~s$^{-1}$\fi}
\newcommand{\ergscm}{\ifmmode {\rm erg~s}^{-1} \else erg~s$^{-1}$ cm$^{-2}$\fi}
\newcommand{\Msun}{\ifmmode M_{\odot} \else $M_{\odot}$\fi }
\newcommand{\Lsun}{\ifmmode {\rm L}_{\odot} \else L$_{\odot}$\fi}
\newcommand{\qo}{\ifmmode q_{\rm o} \else $q_{\rm o}$\fi}
\newcommand{\Ho}{\ifmmode H_{\rm o} \else $H_{\rm o}$\fi}
\newcommand{\ho}{\ifmmode h_{\rm o} \else $h_{\rm o}$\fi}

\newcommand{\vFWHM}{\ifmmode v_{\mbox{\tiny FWHM}} \else
                    $v_{\mbox{\tiny FWHM}}$\fi}
\newcommand{\CCF}{\ifmmode F_{\it CCF} \else $F_{\it CCF}$\fi}
\newcommand{\ACF}{\ifmmode F_{\it ACF} \else $F_{\it ACF}$\fi}
\newcommand{\Halpha}{\ifmmode {\rm H}\alpha \else H$\alpha$\fi}
\newcommand{\Hbeta}{\ifmmode {\rm H}\beta \else H$\beta$\fi}
\newcommand{\Hgamma}{\ifmmode {\rm H}\gamma \else H$\gamma$\fi}
\newcommand{\Hdelta}{\ifmmode {\rm H}\delta \else H$\delta$\fi}
\newcommand{\Lya}{\ifmmode {\rm Ly}\alpha \else Ly$\alpha$\fi}
\newcommand{\Lyb}{\ifmmode {\rm Ly}\beta \else Ly$\beta$\fi}
\newcommand{\HeI}{\ifmmode {\rm He}\,{\sc i}\,\lambda5876 \else 
	          He\,{\sc i}\,$\lambda5876$\fi}
\newcommand{\HeII}{\ifmmode {\rm He}\,{\sc ii}\,\lambda4686 \else 
	           He\,{\sc ii}\,$\lambda4686$\fi}

\newcommand{\heii}{\ifmmode \makebox{{\rm He}\,{\sc ii}} \else He\,{\sc ii}\fi}

\newcommand{\ciii}{\ifmmode {\rm C}\,{\sc iii} \else C\,{\sc iii}\fi}
\newcommand{\civ}{C\,{\sc iv}}

\newcommand{\oiii}{O\,{\sc iii}}

\newcommand{\siiv}{Si\,{\sc iv}}

\def\fake2{\hphantom{3}}

\shorttitle{NGC\,7469}
\shortauthors{Peterson et al.}

\begin{document}

\title{Reverberation Mapping of the Seyfert 1 Galaxy NGC\,7469}
\author{B.~M.~Peterson\altaffilmark{1,2},
C.~J.~Grier\altaffilmark{1,3},
Keith Horne\altaffilmark{4},
R.~W.~Pogge\altaffilmark{1,2},
M.~C.~Bentz\altaffilmark{5},
G.~De~Rosa\altaffilmark{1,6},
K.~D.~Denney\altaffilmark{1,7},
Paul~Martini\altaffilmark{1,2},
S.~G.~Sergeev\altaffilmark{8},
S.~Kaspi\altaffilmark{9,10},
T.~Minezaki\altaffilmark{11},
Y.~Zu\altaffilmark{1,12},
C.~S.~Kochanek\altaffilmark{1,2},
R.J.~Siverd\altaffilmark{13},
B.~Shappee\altaffilmark{1},
C.~Araya~Salvo\altaffilmark{1},
T.~G.~Beatty\altaffilmark{1,3},
J.~C.~Bird\altaffilmark{1,13},
D.~J.~Bord\altaffilmark{14},
G.~A.~Borman\altaffilmark{8},
X.~Che\altaffilmark{15},
C.-T.~Chen\altaffilmark{16},
S.~A.~Cohen\altaffilmark{16},
M.~Dietrich\altaffilmark{17,18},
V.~T.~Doroshenko\altaffilmark{8,19},
T.~Drake\altaffilmark{5},
Yu.~S.~Efimov\altaffilmark{8,*},
N.~Free\altaffilmark{17},
I.~Ginsburg\altaffilmark{16},
C.~B.~Henderson\altaffilmark{1},
A.~L.~King\altaffilmark{15},
S.~Koshida\altaffilmark{11,20},
K.~Mogren\altaffilmark{1},
M.~Molina\altaffilmark{1,3},
A.~M.~Mosquera\altaffilmark{1},
K.~Motohara\altaffilmark{11},
S.~V.~Nazarov\altaffilmark{8},
D.~N.~Okhmat\altaffilmark{8},
O.~Pejcha\altaffilmark{1,21},
S.~Rafter\altaffilmark{10},
J.~C.~Shields\altaffilmark{17},
D.~M.~Skowron\altaffilmark{1,22},
J.~Skowron\altaffilmark{1,22},
M.~Valluri\altaffilmark{15},
J.~L.~van~Saders\altaffilmark{1},
and Y.~Yoshii\altaffilmark{11}
}

\altaffiltext{1}{Department of Astronomy, The Ohio State University,
140 W 18th Ave, Columbus, OH 43210} 
\altaffiltext{2}{Center for
Cosmology \& AstroParticle Physics, The Ohio State University, 191
West Woodruff Ave, Columbus, OH 43210} 
\altaffiltext{3} {Department of Astronomy and Astrophysics,
Eberley College of Science, Penn State University,
525 Davey Laboratory, University Park, PA 16802}
\altaffiltext{4}{SUPA Physics \& Astronomy, University of St. Andrews, Fife, 
KY16 9SS Scotland, UK}
\altaffiltext{5}{Department of Physics and Astronomy, Georgia State
University, Astronomy Offices, 25 Park Place, Suite 605,
Atlanta, GA 30303} 
\altaffiltext{6}{Space Telescope Science Institute, 3700 San Martin
Drive, Baltimore, MD 21218}
\altaffiltext{7}{NSF Postdoctoral Research Fellow}
\altaffiltext{8}{Crimean Astrophysical
Observatory, P/O Nauchny Crimea 298409, Russia}
\altaffiltext{9}{School of Physics and Astronomy, Raymond and Beverly
Sackler Faculty of Exact Sciences, Tel Aviv University, Tel Aviv
69978, Israel} 
\altaffiltext{10}{Physics Department, Technion, Haifa
32000, Israel} 
\altaffiltext{11}{Institute of Astronomy, School of Science, University 
of Tokyo, 2-21-1, Osawa, Mitaka, 181-0015, Tokyo, Japan}
\altaffiltext{12}{McWilliam Center for Cosmology, Department of Physics,
Carnegie Mellon University, 5000 Forbes Ave., Pittsburgh, PA 15213}
\altaffiltext{13}{Department of Physics and Astronomy, Vanderbilt University, 
6301 Stevenson Center, Nashville, TN 37235}
\altaffiltext{14}{Department of Natural Sciences, The
University of Michigan - Dearborn, 4901 Evergreen Rd, Dearborn, MI
48128} 
\altaffiltext{15}{Department of Astronomy, University
of Michigan, 500 Church Street, Ann Arbor, MI 41809}
\altaffiltext{16}{Department of Physics and Astronomy, Dartmouth
College, 6127 Wilder Laboratory, Hanover, NH 03755}
\altaffiltext{17}{Department of Physics \& Astronomy, Ohio
University, Athens, OH 45701} 
\altaffiltext{18}{Department of Physical and Earth Sciences, Worcester
State University, 486 Chandler Street, Worcester, MA 01602}
\altaffiltext{19}{South Station of the Moscow MV Lomonosov State University, 
Moscow, Russia, P/O Nauchny, 298409 Crimea, Russia} 
\altaffiltext{20}{Center for AstroEngineering and Department of Electrical
Engineering, Pontificia Universidad Cat\'{o}lica de Chile,
Av.\ Vicu\~{n}a McKenna 4868, Santiago, Chile}
\altaffiltext{21}{Hubble and Lyman Spitzer Fellow,
Department of Astrophysical Sciences,
Princeton University, 4 Ivy Lane, Peyton Hall, Princeton, NJ 08544}
\altaffiltext{22}{Warsaw University Observatory,
Aleje Ujazdowskie 4,
00-478 Warszawa, Poland}
\altaffiltext{*}{Deceased, 2011 October 21}

\begin{abstract}
A large reverberation mapping study of
the Seyfert 1 galaxy NGC\,7469 has yielded emission-line lags for
\Hbeta\,$\lambda4861$ and \heii\,$\lambda4686$ and a central black hole mass
measurement $M_{\rm BH} \approx 1 \times 10^{7}\,\Msun$,
consistent with previous measurements.
A very low level of variability during the monitoring campaign
precluded meeting our original goal of recovering velocity--delay
maps from the data, but with the new \Hbeta\ measurement,
NGC\,7469 is no longer an outlier in the relationship between
the size of the \Hbeta-emitting broad-line region and the AGN luminosity.
It was necessary to detrend the continuum and  
\Hbeta\ and \heii\,$\lambda4686$ line light curves 
and those from archival UV data for different time-series
analysis methods to yield consistent results.
\end{abstract}

\keywords{galaxies: active ---
galaxies: individual (NGC\,7469) ---
galaxies: nuclei ---
galaxies: Seyfert
}


\section{INTRODUCTION}
\label{section:intro}

\setcounter{footnote}{0}

Reverberation mapping \citep{Blandford82,Peterson93,Peterson13} 
is a standard tool for probing the structure and kinematics of 
the broad-line region (BLR) in active galactic nuclei
(AGNs). In its simplest form, the mean time delay between continuum
and emission-line variations is measured, typically by cross-correlation
of the respective light curves, and it is assumed that this represents the 
mean light-travel time across the BLR. By combining this with the emission-line
width, which is assumed to reflect the velocity dispersion of gas whose
motions are dominated by the mass of the central black hole, the black 
hole mass can be determined. Reverberation mapping in this form has
been used to measure the black hole masses in nearly 50 AGNs
\citep[for a recent compilation, see][]{Bentz13} to a typical
accuracy of $\sim 0.4$\,dex.

A decade ago, we undertook a consistent reanalysis of the
reverberation-mapping data base that existed at that time
\citep{Peterson04}.  In the course of this work, we identified a
number of cases where the BLR radius, mass estimates, or both would
clearly benefit from improved monitoring. NGC 7469, one of Seyfert's
(1943) original galaxies distinguished by an abnormally bright core,
is one such example.

The Seyfert 1 galaxy NGC 7469 was the subject of a large coordinated
X-ray \citep{Nandra98,Nandra00}, ultraviolet (UV; \citealt{Wanders97}), 
and optical \citep{Collier98}
monitoring campaign in 1996 June -- July with
the {\em Rossi X-Ray Timing Explorer (RXTE)},
the {\em International Ultraviolet Explorer (IUE)}, 
and several ground-based telescopes by the International AGN
Watch consortium \citep{Alloin94}. The program was supplemented
with very high sampling-rate spectrophotometry for
10 hours with the Faint Object Spectrograph (FOS) on
{\em Hubble Space Telescope} ({\em HST}) to look for
very short timescale continuum variability \citep{Welsh98}.
This was the final year of {\em IUE}
operations and an intensive reverberation-mapping program
was intended to be part of a ``Grand Finale.'' One of the primary
science goals of the {\em IUE} program
was to recover a velocity--delay map 
\citep{Horne04} for the
strong UV emission lines. The original target
for the campaign was the Seyfert 1 galaxy Mrk 335, but
an unfortunate gyroscope failure on {\em IUE} in 1996 March forced selection of
an alternative target. NGC 7469 was deemed to be the most promising
of a very small number of AGNs that would be accessible 
during the time period allocated for this program.
Unfortunately, managing spacecraft pointing with {\em IUE} remained
an issue that compromised the quality of the resulting data. The duration
of the intensive part of the monitoring program was very limited and as a 
result, while reverberation lags were measured, it was not possible to
obtain a reliable velocity--delay map from the {\em IUE} spectra.

The high-sampling rate UV/optical light curves did, 
however, reveal for the first time
a statistically significant lag between continuum variations in
the UV and those following at longer wavelengths
\citep{Wanders97,Collier98}.
The variations at 1825\,\AA, 4845\,\AA, and 6962\,\AA\ follow
those at 1315\,\AA\ by $0.22^{+0.12}_{-0.13}$,
$1.25^{+0.48}_{-0.35}$, and
$1.84^{+0.93}_{-0.94}$ days, respectively \citep{Peterson98}.
Similar, but lower significance, interband continuum lags have
also been seen in NGC 4151 \citep{Peterson98} and
interband lags in the optical alone have been detected 
in 14 AGNs by \cite{Sergeev05} at varying levels of significance.
More recently, multiwavelength monitoring of NGC 2617, which
recently underwent a dramatic change in ``type'' from Seyfert 1.8 to
Seyfert 1, revealed that flux variations in all the continuum bands
from the UV to NIR follow variations in the X-ray, with the lag increasing
with wavelength. Moreover, the structure of the light curves becomes 
smoother with increasing wavelength, indicating some 
``time-smearing'' associated with continuum reprocessing \citep{Shappee14}.

Detection of interband lags is important, as it points to the mechanisms 
that cause continuum variability. There are also important implications
for reverberation mapping. Specifically, it is necessarily assumed that
the observable optical continuum is a reasonable proxy for the
unobservable UV continuum that photoionizes the broad-line gas and 
drives the emission-line variations. A small time delay between the 
variations in the ionizing continuum and the optical continuum will result
in a small underestimate of the BLR size. 
Even more important, however, is that 
if the optical continuum is a smoothed or time-smeared version of the
ionizing continuum, there might be structure in the emission-line
light curves that may not be present in the optical continuum light curves
and recovery of the detailed structure of the BLR becomes
more difficult.

The relationship between the continuum variations $\Delta C(t)$
and velocity-resolved emission-line variations $\Delta L(V,t)$ is
usually expressed mathematically as
\begin{equation}
\label{eq:trans}
\Delta L(V,t) = \int_{0}^{\infty} \Psi (V,\tau) \Delta C(t-\tau )d\tau,
\end{equation}
where $\Psi(V,\tau)$ is the ``transfer function'' \citep{Blandford82},
or velocity--delay map \citep{Horne04}, which is the
observed emission-line response to a delta-function continuum outburst.
This simple linear formulation
is justified by the fact that the continuum and emission-line variations
are generally quite small (10--20\%) on reverberation time scales.
The technical goal of a reverberation program is to recover the
velocity--delay map $\Psi(\Delta V,\tau)$ from the data and thus infer the geometry and
kinematics of the BLR. However, if
the optical continuum light curve is not a good surrogate
for the variability of the ionizing continuum, then the fidelity with which
we can recover velocity--delay maps is fundamentally limited.
On the basis of the data obtained in the observing campaign described
here, we suggest that this may be the case
in NGC 7469.

Here we describe an optical reverberation-mapping monitoring
program on NGC 7469 that was undertaken with the primary goal of
obtaining a velocity--delay map for its \Hbeta\,$\lambda4861$ and 
He\,{\sc ii}\,$\lambda4686$ emission lines.
We describe the observations and data analysis
in \S{2}. Our time-series analysis is presented in \S{3} and
our black hole mass measurement is explained in \S{4}.
We briefly discuss and summarize our results in \S{5}.

\section{OBSERVATIONS \& DATA ANALYSIS}
The data used in this study were obtained during a four-month long
observing campaign carried out in late 2010. The results
for the other five objects observed in this campaign have been
published by \cite{Grier12b}. We follow the data analysis procedures
described in that study.  When needed, we adopt a
cosmological model with $\Omega_{m} = 0.3$, $\Omega_{\Lambda} = 0.70$, and
$H_0 = 70$ km sec$^{-1}$ Mpc$^{-1}$.

\subsection{Observations}
We obtained 73 spectra with the Boller and Chivens CCD spectrograph
on the MDM Observatory 1.3-m McGraw-Hill telescope on Kitt Peak. The
data were collected over the course of 120 nights from 2010 August 31
to December 28. We used a 350 mm$^{-1}$ grating to obtain a
dispersion of 1.33\,\AA\,pixel$^{-1}$. We set the grating for a
central wavelength of 5150\,\AA, which resulted in spectral coverage
over the range 4400\,\AA \ to 5850\,\AA. The slit was oriented
North--South (position angle ${\rm PA} = 0^{\rm o}$) with a 
projected width of $5\farcs0$ that produces
a spectral resolution of 7.9\,\AA. We used an extraction
window of $12\farcs0$ along the slit.

To supplement our spectra in estimating the driving continuum light
curve, we obtained $V$-band imaging observations of NGC\,7469 at
several additional observatories. We obtained 74 images using the 70-cm
telescope at the Crimean Astrophysical Observatory (CrAO) with the
AP7p CCD, which has 512 $\times$ 512 pixels with a 15\arcmin
$\times$15\arcmin \ field of view when mounted at prime focus. We also
obtained 66 epochs from the 46-cm Centurion telescope at Wise
Observatory of Tel-Aviv University using an STL-6303E CCD with 3072
$\times$ 2048 pixels, with a field of view of
75\arcmin$\times$50\arcmin. Further $V$-band
observations were obtained for 7 epochs using the University of
Tokyo's 1.0-m miniTAO telescope stationed in Chile. We used the ANIR
CCD camera \citep{Motohara08}, which has a pixel scale of 
$0\farcs34\,{\rm pixel}^{-1}$ and a
field of view of $6'\times 6'$. Finally, we obtained
observations of NGC\,7469 for 56 epochs using the SMARTS CTIO 1.3-m
telescope in Chile with the ANDICAM CCD camera, which has a field of
view of $6' \times 6'$, 1024$\times$1024 pixels, and a
pixel scale of $0\farcs371\, {\rm pixel}^{-1}$.

\subsection{Data Processing and Light Curves}
\begin{figure}
\begin{center}
\epsscale{1.0}
\plotone{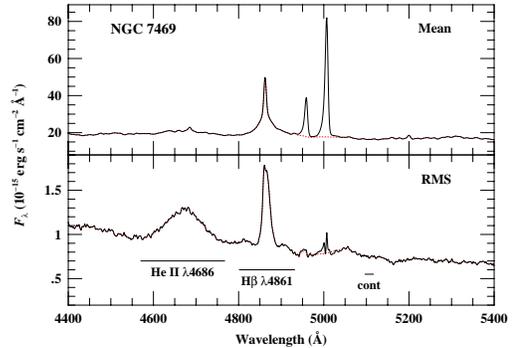}
\caption{Rest-frame ($z=0.01632$) mean (top panel) and RMS (bottom panel)
spectra of NGC\,7469. The dashed red lines show the spectra where
the [\oiii]\,$\lambda\lambda4959$, 5007 lines are removed
prior to combining the individual spectra. The integration limits
for \heii\,$\lambda4686$, \Hbeta\,$\lambda4861$, and the continuum
are indicated by horizontal lines in the bottom panel.} 
\label{fig:meanrms}
\end{center}
\end{figure}

To place the reduced spectra on an absolute flux scale, we assume that
the [O\,{\sc iii}]\,$\lambda5007$ narrow-line flux is constant. 
Because of the relatively long light-travel time across the 
narrow-line region and because of the long recombination times,
this is a good assumption on reverberation timescales, although
narrow-line variability has been detected 
in other AGNs on timescales as short as
years \citep[e.g.,][]{PetersonPlus13}. We
used a reference spectrum created by averaging spectra taken on
photometric nights and scale all our spectra to match this
reference spectrum. We measure the average 
[O\,{\sc iii}]\,$\lambda5007$ flux in the reference
spectrum to be $(6.14 \pm 0.12) \times 10^{-13}$\,\ergscm, 
which we adopt as the absolute
flux for this object. This value is in general agreement with
the [O\,{\sc iii}]\,$\lambda5007$ flux
reported by \cite{Collier98}. We then scaled each individual spectrum
to the reference spectrum using a $\chi^2$ goodness-of-fit estimator
method to minimize the flux differences between the spectra
(\citealt{vanGroningen92}). Figure~\ref{fig:meanrms} shows the mean
and root mean square residual (RMS) spectra of NGC\,7469 based on the
calibrated MDM spectra. Emission-line light curves were created by
fitting a linear continuum underneath the emission lines in each
scaled spectrum and integrating the flux above them. The \Hbeta$\,\lambda4861$
integrations were done between the observed-frame wavelengths of
4880--5012 \AA, and the \heii$\,\lambda4686$ fluxes were measured between
4645--4845 \AA. The 5100\,\AA \ continuum light curves were created by
taking the average flux measured from 5180--5200\,\AA \ in the
observed frame.

\begin{deluxetable}{lcc}
\tablewidth{0pt}
\tablecaption{Continuum Fluxes\tablenotemark{*}}
\tablehead{
\colhead{HJD\tablenotemark{a}} &
\colhead{Observatory\tablenotemark{b}} & 
\colhead{$F_{\rm cont}(5100\,{\rm \AA})$\tablenotemark{c}} 
}
\startdata
5430.430 & C& $14.90 \pm 0.17$	 \\
5430.780 & S& $14.88 \pm 0.02$	 \\
5431.420 & C& $14.66 \pm 0.15$	 \\
5432.410 & C& $15.13 \pm 0.16$	 \\
5433.370 & C& $15.36 \pm 0.25$	 \\
5436.390 & C& $15.34 \pm 0.15$	 \\
5437.340 & W& $15.43 \pm 0.02$	 \\
5437.400 & C& $15.75 \pm 0.13$	 \\
5437.790 & S& $15.53 \pm 0.01$	 \\
5438.360 & C& $15.51 \pm 0.21$	 \\
5438.390 & W& $15.46 \pm 0.02$	 \\
5438.810 & S& $15.46 \pm 0.01$	 \\
\enddata %
\tablenotetext{*}{Table is given in full in the published version.}
\tablenotetext{a}{Heliocentric Julian Date ($-2450000$).}
\tablenotetext{b}{Observatory Code: C=CRAO, W= WISE, M=MDM, S=SMARTS, T=miniTAO}
\tablenotetext{c}{Continuum fluxes are in units of $10^{-15}$ ergs s$^{-1}$ cm$^{-2}$ \AA$^{-1}$.}
\label{Table:confluxes} 
\end{deluxetable}

We produced light curves from our $V$-band photometry using the image
subtraction software package {\tt ISIS} (\citealt{Alard98};
\citealt{Alard00}). We follow the procedures of \cite{Shappee11},
wherein the images are first aligned using the program {\tt Sexterp}
\citep{Siverd12} including its optional resampling utility {\tt is3\_interp}.
We then follow the steps outlined by Alard, using {\tt ISIS} to
create a reference image for the field using the 20--30 images with
the best seeing and lowest background counts. {\tt ISIS} convolves the
images with a spatially variable convolution kernel to transform all
images to the same point-spread function (PSF) and background
level. The resulting images are stacked using a 3$\sigma$ rejection
limit from the median. We then used {\tt ISIS} to convolve the reference
image to match each individual image in the data set and subtract each
individual frame from the convolved reference image. We
extract light curves for the nucleus of the galaxy from these
subtracted images using ISIS to place a PSF-weighted aperture over the
nucleus and measure the residual flux.

\begin{figure}
\begin{center}
\epsscale{1.0}
\plotone{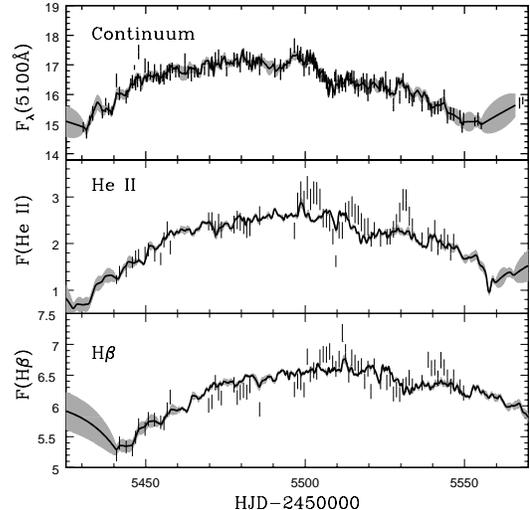}
\caption{Combined light curves for NGC\,7469, along with the {\tt JAVELIN}
models. The black vertical bars show the original data,
and the solid black line shows the mean of the {\tt JAVELIN} light curve
models {\bf most} consistent with the data. 
The gray shaded region shows the standard
deviation of values about the mean. Continuum fluxes are in units of
$10^{-15}$ erg s$^{-1}$ cm$^{-2}$ \AA$^{-1}$ and emission-line fluxes
are in units of $10^{-13}$ erg s$^{-1}$ cm$^{-2}$.}
\label{fig:lcs}
\end{center}
\end{figure}

The spectroscopic continuum light curve was then merged with the
photometric light curves to create our final continuum light curve. To
correct for the differences in host-galaxy starlight that enters the
apertures, we applied a multiplicative scale factor as well as an
additive flux adjustment to each photometric light curve (see
\citealt{Peterson95}). The merged continuum light curve and the MDM
spectroscopic light curves for \Hbeta \ and \heii \ are shown as the
black vertical bars in Figure~\ref{fig:lcs}. The continuum light curve fluxes
are given in Table~\ref{Table:confluxes}, with each data point labeled
according to the observatory at which it was obtained. The \Hbeta \
and \heii \ fluxes from the MDM spectra are listed in
Table~\ref{Table:linefluxes}. Final light curve statistics for all
three light curves are given in Table~\ref{Table:lcstats}. We also include
the statistics for the AGN continuum only, with our best estimate of
the starlight contamination ($F_{\rm gal} = (8.7 \pm 0.9) \times
10^{-15}$ ergs s$^{-1}$ cm$^{-2}$ \AA$^{-1}$, \citealt{Bentz13}\footnote{The entry
for the host galaxy flux for NGC\,7469 in Table 12 of
\cite{Bentz13} is in error. The correct value is used here.}) subtracted from
each of the continuum measurements 
that are given in Table~\ref{Table:confluxes}.
It is worth noting that the mean optical flux from the AGN alone 
during the AGN Watch program was
$F_{\rm AGN} =5.14 \times 10^{-15}$\,erg\,s$^{-1}$\,cm$^{-2}$\,\AA$^{-1}$
\citep{Bentz09};
i.e., the AGN was $\sim 52\%$ more luminous in 2010 than it was in 1996.

\begin{deluxetable}{lcc}
\tablewidth{0pt}
\tablecaption{Emission-Line Fluxes\tablenotemark{*}}
\tablehead{
\colhead{HJD\tablenotemark{a}} & 
\colhead{$F(\Hbeta)$\tablenotemark{b}} & 
\colhead{$F(\mbox{\rm He\,{\sc ii}}\,\lambda4686)$\tablenotemark{b}} 
}
\startdata
 5440.801 & $5.203 \pm 0.102$ &$ 1.228 \pm 0.096$ \\
 5441.763 & $5.398 \pm 0.106$ &$ 1.413 \pm 0.111$ \\
 5443.820 & $5.331 \pm 0.104$ &$ 1.391 \pm 0.109$ \\
 5445.779 & $5.286 \pm 0.104$ &$ 1.596 \pm 0.125$ \\
 5446.787 & $5.491 \pm 0.108$ &$ 1.719 \pm 0.135$ \\
 5447.767 & $5.688 \pm 0.111$ &$ 1.748 \pm 0.137$ \\
 5449.767 & $5.624 \pm 0.110$ &$ 1.790 \pm 0.140$ \\
 5450.758 & $5.788 \pm 0.113$ &$ 1.871 \pm 0.147$ \\
 5452.773 & $5.778 \pm 0.113$ &$ 1.751 \pm 0.137$ \\
 5454.698 & $5.734 \pm 0.112$ &$ 1.671 \pm 0.131$ \\
 5455.696 & $5.717 \pm 0.112$ &$ 2.080 \pm 0.163$ \\
 5456.744 & $5.943 \pm 0.116$ &$ 2.295 \pm 0.180$ \\
 5457.695 & $6.145 \pm 0.120$ &$ 1.918 \pm 0.150$ \\
\enddata     	    
\tablenotetext{*}{Table is given in full in the published version.}
\tablenotetext{a}{Heliocentric Julian Date ($-2450000$).}
\tablenotetext{b}{Emission-line fluxes are in units of $10^{-13}$ ergs s$^{-1}$ cm$^{-2}$.}
\label{Table:linefluxes} 
\end{deluxetable}

\section {TIME-SERIES ANALYSIS}

\begin{figure}
\begin{center}
\epsscale{1.0}
\plotone{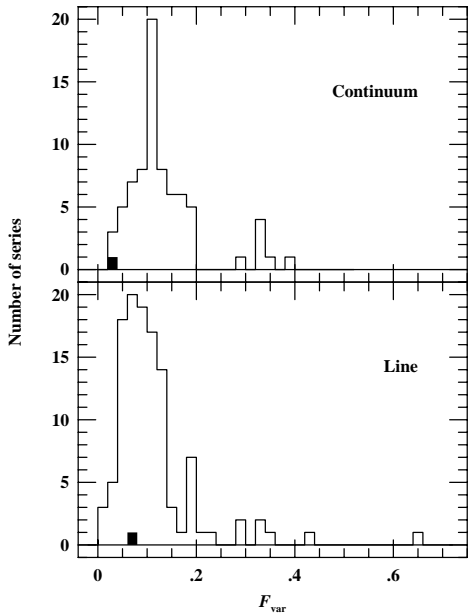}
\caption{Distribution of fractional variability measures
$F_{\rm var}$ for successful reverberation-mapping campaigns compiled
by \cite{Peterson04}. The upper panel shows the continuum
variations, usually in the optical and uncorrected for
starlight contamination, and the lower panel shows
line variations, usually for \Hbeta\ and uncorrected for
narrow-line contamination. The values for the current campaign
on NGC 7469, shown in black, indicate how comparatively little
variability was detected in this campaign.}
\label{fig:fvar}
\end{center}
\end{figure}

Inspection of the light curves
in Figure~\ref{fig:lcs} and the variability statistics 
(Table~\ref{Table:lcstats}) show that the overall levels of
variability were of much lower amplitude than normally desirable for
a reverberation experiment. Figure~\ref{fig:fvar} shows
a comparison of the fractional variation $F_{\rm var}$ observed
over this campaign with 116 previous successful reverberation
time series \citep{Peterson04}. The continuum light curve lacks the
strong short-timescale variations that produce the
clearest reverberation signatures.
This necessarily severely limits the
amount of information that we can extract from these data.
Also, we note that there are structures in
the emission-line light curves that are not present in the 
continuum light curve. As mentioned in \S{\ref{section:intro}},
this leads us to suspect that the far-UV ionizing continuum light curve
has more short-timescale structure than the reprocessed optical
continuum. The broad-line gas
reprocesses the ionizing photons into emission-line photons
rapidly (as the recombination time at BLR densities is less than
an hour). The continuum reprocessing timescale, on the other hand,
must be somewhat longer and thus slightly smears out the shorter 
timescale variations in the shorter-wavelength continuum.
We keep this in mind as we consider the response of the
emission lines to the continuum variations.

\begin{deluxetable*}{lccccccc}
\tablewidth{0pt}
\tablecaption{Light Curve Statistics}
\tablehead{
\colhead{ } &
\colhead{ } &
\multicolumn{2}{c}{Sampling} &
\colhead{ } &
\colhead{Mean} \\
\colhead{Time} &
\colhead{ } &
\multicolumn{2}{c}{Interval (days)} &
\colhead{Mean} &
\colhead{Fractional} \\
\colhead{Series } &
\colhead{$N$} &
\colhead{$\langle T \rangle$} &
\colhead{$T_{\rm median}$} &
\colhead{Flux} &
\colhead{Error} &
\colhead{$F_{\rm var}$} &
\colhead{$R_{\rm max}$} \\
\colhead{(1)} &
\colhead{(2)} &
\colhead{(3)} &
\colhead{(4)} &
\colhead{(5)} &
\colhead{(6)} &
\colhead{(7)} &
\colhead{(8)} 
} 
\startdata
5100\,\AA\         & 276 & 0.5 & 0.40  & $16.54\pm0.60$  & 0.007 & 0.035 & $1.20\pm 0.02$\\
5100\,\AA, AGN only& 276 & 0.5 & 0.40  & $7.84\pm0.60$  & 0.015 & 0.074 & $1.50\pm0.06$\\
$\Hbeta\,\lambda4861$& 73  & 1.5 & 1.00  & $6.31\pm0.41$  & 0.020 & 0.062 & $1.38\pm 0.04$\\
\heii\,$\lambda4686$& 73  & 1.5 & 1.00  & $2.29\pm0.44$  & 0.078 & 0.172 & $2.60\pm 0.29$
\enddata
\tablenotetext{*}{Column (1) lists
the spectral feature, and column (2) gives the number of points in the
individual light curves. Columns (3) and (4) list the average and
median time spacing between observations, respectively. Column (5)
gives the mean flux of the feature in the observed frame, and column
(6) shows the mean fractional error that is computed based on
observations that are closely spaced in time. Column (7) gives the
excess variance, defined by
\begin{equation}
F_{\rm var} = \frac{\sqrt{\sigma^{2}-\delta{^2}}}{\langle f\rangle}
\end{equation}
where $\sigma^{2}$ is the flux variance of the observations,
$\delta^{2}$ is the mean square uncertainty, and $\langle f \rangle$
is the mean observed flux. Column (8) is the ratio of the maximum to
minimum flux in each light curve.}
\tablenotetext{*}{Continuum and emission-line fluxes 
are given in $10^{-15}$ \ergscm\AA$^{-1}$ and $10^{-13}$ \ergscm, 
respectively.}
\label{Table:lcstats}
\end{deluxetable*}

We use two different methods to examine the 
time-delayed response of the 
\Hbeta\,$\lambda4861$ and 
\heii$\,\lambda4686$ emission lines to the continuum variations, 
as we describe below. We also attempted to recover velocity--delay maps
as we did for other sources observed in the same campaign
\citep{Grier13a}, but we were unsuccessful on account of the
low level of variability in this source during the monitoring campaign.

\subsection{Cross-Correlation Analysis} 
\begin{figure}
\begin{center}
\epsscale{1.0}
\plotone{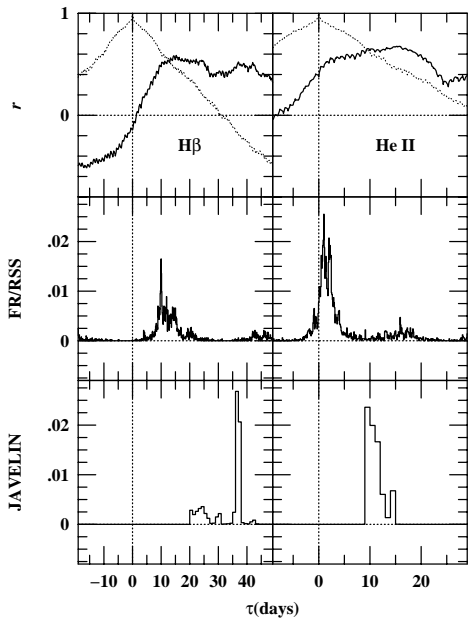}
\caption{Time-series analysis results for \Hbeta\ (left column)
and \heii\ (right column), based on the light curves given in
Tables~\ref{Table:confluxes} and \ref{Table:linefluxes} and
shown in Figure~\ref{fig:lcs}.
The top row shows the cross-correlation functions; the 
solid black lines show the CCFs for
each emission line, and the gray dashed lines show the continuum
autocorrelation function.
The middle row shows the 
cross-correlation centroid distribution from 1000 FR/RSS
realizations. The bottom row shows the posterior
lag distributions from {\tt JAVELIN}. The cross-correlation functions
have no clearly defined maxima and the {\tt JAVELIN}
and cross-correlation results are in poor agreement.}
\label{fig:timeseries}
\end{center}
\end{figure}

For an initial attempt to determine the emission-line lags, we 
cross-correlated the continuum light curve (Table~\ref{Table:confluxes})
with the emission-line light curves (Table~\ref{Table:linefluxes}).
The methodology we use was first described by
\cite{Gaskell86} and
\cite{Gaskell87} and later significantly modified by \cite{White94} 
and updated by \cite{Peterson98} and \cite{Peterson04}.
The cross-correlation functions (CCFs) are shown in top panels of
Figure~\ref{fig:timeseries}, along with the continuum
autocorrelation function (ACF).
The cross-correlation results are given in the first two
rows of Table~\ref{Table:lags}, 
where $\tau_{\rm peak}$ is the value of the time delay
or lag where the CCF is maximized ($r_{\rm max}$).
The centroid of the CCF peak $\tau_{\rm cent}$ is computed from
all neighboring points near $\tau_{\rm peak}$ with $r(\tau) \geq 0.8 r_{\rm max}$,
although some experimentation shows that the centroid is
insensitive to the threshold used in the computation.
The quoted 1$\sigma$ uncertainties were determined by using the 
model-independent Monte Carlo method of 
flux randomization and random subset sampling (FR/RSS) 
described by \cite{Peterson98} and \cite{Peterson04}. The
cross-correlation centroid distributions from this
process are also shown in the middle panels of Figure~\ref{fig:timeseries}. 

\begin{deluxetable}{lccc} 
\tablewidth{0pt} 
\tablecaption{\Hbeta, \HeII, and \civ\ Time Series Results$^{*}$} 
\tablehead{ 
\colhead{} &
\multicolumn{2}{c}{\underline{This Campaign}} & 
\colhead{\underline{Archival}} \\
\colhead{} &
\colhead{\Hbeta} & 
\colhead{\heii} &
\colhead{\civ} \\
\colhead{Parameter} &
\colhead{(days)} & 
\colhead{(days)} & 
\colhead{(days)} \\
\colhead{(1)} & 
\colhead{(2)} & 
\colhead{(3)} &
\colhead{(4)}  
} 
\startdata
{\bf Original:} \\
$\tau_{\rm cent}$      & $11.8^{+4.2}_{-2.6}$  & $1.9^{+7.5}_{-1.6}$ & $2.6^{+0.3}_{-0.3}$ \\
$\tau_{\rm peak} $     & $11.8^{+4.2}_{-2.7}$  & $2.0^{+7.4}_{-1.7}$ & $2.5^{+0.5}_{-0.5}$ \\
$\Delta\tau_{\rm cent}(\makebox{\Hbeta\ $-$ He\,{\sc ii}})$ 
                       & $8.4^{+2.0}_{-2.0}$    & $\ldots$ & $\ldots$\\ 
$\tau_{\rm {\tt JAVELIN}}$  & $23.6^{+1.8}_{-2.7}$  & $10.4^{+2.1}_{-0.7}$&$10.8^{+0.2}_{-0.2}$ \\
{\bf Detrended:}\\
$\tau_{\rm cent}$      & $10.9^{+3.5}_{-1.3}$    & $1.3^{+0.9}_{-0.7} $& $2.3^{+0.3}_{-0.3}$ \\
$\tau_{\rm peak} $     & $11.2^{+3.3}_{-1.4}$   & $1.2^{+1.1}_{-0.8}$ & $2.2^{+0.5}_{-0.3}$ \\
$\Delta\tau_{\rm cent}(\makebox{\Hbeta\ $-$ He\,{\sc ii}})$ 
                       & $9.0^{+2.0}_{-1.5}$    & $\ldots$ & $\ldots$\\ 
$\tau_{\rm {\tt JAVELIN}}$  & $10.0^{+1.2}_{-0.4}$  & $0.8^{+0.7}_{-0.2}$ & $2.3^{+0.3}_{-0.2}$
\enddata 
\tablenotetext{*}{All time delays are given in the observed frame.
The ``original'' \civ\ time delays are from \cite{Peterson04} and
\cite{Zu11}.}
\label{Table:lags}
\end{deluxetable}

The CCFs for both lines (Figure~\ref{fig:timeseries}) have broad plateaus
extending from $\sim 10$ days for \Hbeta\ and from
close to zero days for \heii\ to much larger lags, leaving the correct
lags uncertain, although the \Hbeta\ lag is clearly longer than the 4--5\,day
lag from the 1996 campaign \citep{Collier98,Peterson04}.
The FR/RSS centroid distribution functions shown in the
middle panels of Figure~\ref{fig:timeseries} are less well-defined 
than in most cases, and the \heii\ centroid distribution function
has a broad tail extending to nearly 20 days.

\subsection{{\tt JAVELIN} Analysis}
\cite{Zu11} have developed an alternative method of measuring
reverberation time lags called Stochastic Process Estimation for AGN
Reverberation ({\tt SPEAR}), that was subsequently upgraded to
the software package we used in our analysis, 
{\tt JAVELIN}\footnote[1]{
\url{http://www.astronomy.ohio-state.edu/$\sim$yingzu/codes.html\#javelin}}.
{\tt SPEAR} and
{\tt JAVELIN} have been used successfully to determine time lags by
\cite{Grier12a, Grier12b}, and to model continuum light curve
behavior \citep{Grier13a}. As with cross correlation, this method
assumes all emission-line light curves are scaled and shifted versions
of the continuum light curve. {\tt JAVELIN} models the continuum as an
autoregressive process using a damped random walk  model, which
has been demonstrated to be a good statistical model of AGN
variability (e.g., \citealt{Kelly09};
\citealt{Kozlowski10}; \citealt{MacLeod10}; \citealt{MacLeod12};
\citealt{Zu13}). The software explicitly builds a model of the light
curve and transfer function and fits it to the data by maximizing the
likelihood of the model. {\tt JAVELIN} then computes
uncertainties using the Bayesian Markov Chain Monte Carlo method.
 
We used {\tt JAVELIN} to determine the time lag between the 5100 \AA \
continuum and both the \Hbeta \ and \heii \ emission lines. 
The {\tt JAVELIN} results are also given in Table~\ref{Table:lags}.
The posterior distributions of the successful  {\tt JAVELIN} models are shown in 
the bottom panels of Figure~\ref{fig:timeseries}. 
We see that the {\tt JAVELIN} distributions are highly
inconsistent with the FR/RSS distributions immediate above.

\subsection{Analysis of Detrended Light Curves}

When there is little short-timescale variability and much of the variability is
on timescales comparable to the duration of the observing campaign,
aliasing becomes an increasing problem.

The time delays measured by cross correlation and with {\tt JAVELIN}
are much more inconsistent than we usually find in reverberation
studies, almost certainly as a consequence of the low amplitude of
variability. Moreover, what little variability there is seems to be
dominated by long-term quasi-parabolic trends where the light curves
initially slowly rise, then fall.  Trends longer than reverberation
timescales can yield misleading reverberation results as shown by,
e.g., \cite{Grier08}.  \cite{Welsh99} suggested that reverberation
measurements based on cross-correlation analysis could be improved by
``detrending'' the light curves: when light curves are dominated by
trends longer than the reverberation timescale, we fit the light
curves with low-order polynomials, and subtract off these longer-term
trends prior to applying the cross-correlation analysis. In previous
experiments \citep{Denney10}, we found that detrending led to marked
improvement in the results.

\begin{figure}
\begin{center}
\epsscale{1.0}
\plotone{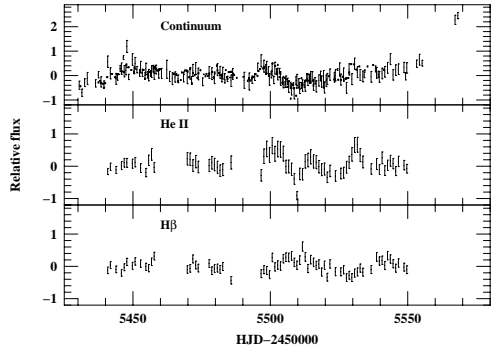}
\caption{Continuum (top), \heii\ (middle), and \Hbeta\ (bottom) light curves after
detrending by subtracting a low-order polynomial from the original
light curves shown in Figure~\ref{fig:lcs}.}
\label{fig:dtlc}
\end{center}
\end{figure}

\begin{figure}
\begin{center}
\epsscale{1.0}
\plotone{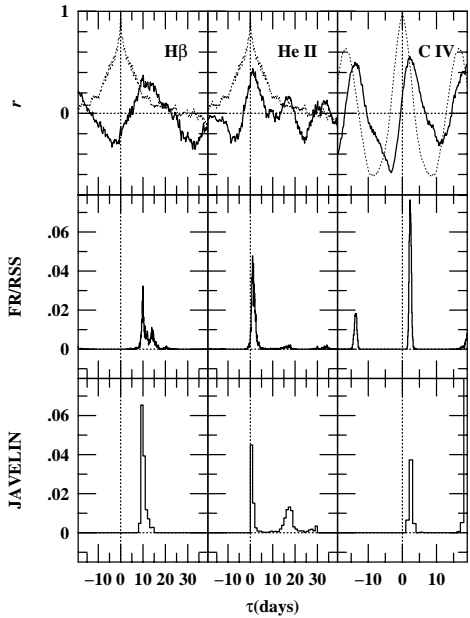}
\caption{Time-series analysis results for \Hbeta\ (left column),
and \heii\ (middle column) based on the detrended light curves as shown in
Figure~\ref{fig:dtlc} and for \civ\,$\lambda1549$ (right column) based on 
archival {\em IUE} data.
The top row shows the cross-correlation functions; the 
solid black lines show the CCFs for
each emission line, and the gray dashed lines show the continuum
autocorrelation function.
The middle row shows the 
cross-correlation centroid distribution from 1000 FR/RSS
realizations. The bottom row shows the 
lag distribution from {\tt JAVELIN}.
The results for \Hbeta\ and \heii\ should be compared with
those shown in Figure~\ref{fig:timeseries}. The results for
the detrended light curves are far more consistent.}
\label{fig:dttimeseries}
\end{center}
\end{figure}

In Figure~\ref{fig:dtlc}, we show the light curves from
Figure~\ref{fig:lcs} after detrending. The cross-correlation results
based on the detrended data are shown in top panels
Figure~\ref{fig:dttimeseries} and listed in Table~\ref{Table:lags},
and the cross-correlation centroid distributions are shown in in the
middle panels of Figure~\ref{fig:dttimeseries}. The posterior lag
distributions from {\tt JAVELIN} are shown in the bottom panels
Figure~\ref{fig:dttimeseries}. Clearly, the He\,{\sc ii} and \Hbeta\
lags are much better defined using the detrended light curves and
there is consistency between the cross-correlation and {\tt JAVELIN}
results. Moreover, direct cross-correlation of the emission lines with
each other yields lags that are statistically indistinguishable for
the original and detrended data ($\Delta \tau_{\rm cent}(\Hbeta -
\heii)$ in Table~\ref{Table:lags}).  The difference is also consistent
with the difference between the two continuum--emission-line lags.

There was a similar discrepancy for the \civ\,$\lambda1549$
emission line in the AGN Watch data
on NGC 7469 \citep{Zu11}. The lags for \heii\,$\lambda 1640$ and
\siiv\,$\lambda1400$ were consistent, but
in the case of the \civ\,$\lambda1549$ line, 
the cross-correlation lag was measured to be $\sim 2.5$ days, while
the {\tt JAVELIN} time delay was nearly 11 days. We
suspected that the \civ\ results might also be improved by detrending. This
did indeed prove to be the case, as {\tt JAVELIN} yields a \civ\ time delay
that is in much better agreement with the cross-correlation result
(Table~\ref{Table:lags} and Figure~\ref{fig:dttimeseries})
 after a simple linear detrending.

\section {LINE WIDTH AND BLACK HOLE MASS CALCULATION}

Assuming that the motion of the BLR gas is dominated by gravity
and that radiation pressure can be neglected, the mass of the central
black hole is given by
\begin{equation}
\label{eq:BHmass}
M_{\rm BH} = f \left( \frac{c \tau \Delta V^2}{G} \right),
\end{equation}
where $\tau$ is the emission-line time delay, $\Delta V$ is the velocity
width, and $f$ is a dimensionless factor that depends
on the structure, kinematics and orientation of the BLR. The quantity
in parentheses in equation (\ref{eq:BHmass}) contains just the observables
and is sometimes referred to as the ``virial product'' $M_{\rm vir}$
(i.e., $M_{\rm BH} = f M_{\rm vir}$).

The BLR velocity dispersion can be characterized by either 
the FWHM or the line dispersion $\sigma_{\rm line}$.
To determine the best value of
the line width and its uncertainty, we use Monte Carlo simulations
similar to those used when determining the lag from the CCF. We run
200 simulations in which we create a mean and RMS spectrum
from a randomly chosen subset of the spectra, obtaining a distribution
of resolution-corrected line widths. We adopt the mean values of FWHM
and $\sigma_{\rm line}$ from these simulations and adopt their standard
deviation as our formal uncertainty. We measure $\sigma_{\rm line}$
and FWHM in both the mean and RMS spectra, and these
appear in Table~\ref{Table:vwidths}. There is some evidence
that $\sigma_{\rm line}$ produces less biased mass measurements than
FWHM (\citealt{Peterson11}), so we prefer to use $\sigma_{\rm line}$
to compute $M_{\rm BH}$. We also prefer to use measurements from the RMS
spectrum, as this eliminates contamination from constant
narrow-line and other slowly varying
components to isolate the broad emission components that
are actually responding to the continuum variations. This
prescription yields the virial products listed in 
Table~\ref{Table:virial}.
\begin{deluxetable}{lccc} 
\tablewidth{0pt} 
\tablecaption{Emission-Line Widths\tablenotemark{a}} 
\tablehead{ 
\colhead{Parameter} & 
\colhead{\Hbeta} & 
\colhead{\heii}  &
\colhead{\civ} \\ 
\colhead{ } &
\colhead{(km\,s$^{-1}$)} &
\colhead{(km\,s$^{-1}$)} & 
\colhead{(km\,s$^{-1}$)} \\
\colhead{(1)} & 
\colhead{(2)} & 
\colhead{(3)} & 
\colhead{(4)}
}
\startdata
$\sigma_{\rm line}$ (mean) 
& $1095\pm 5$ & $2306 \pm 8$ & $1707 \pm 20$ \\
FWHM (mean) 
& $4369 \pm 6$ & $2197 \pm 339$ & $1722 \pm 30$ \\
$\sigma_{\rm line}$ (RMS) 
& $1274 \pm 126$ & $2271 \pm 77$ & $2619 \pm 118$ \\
FWHM (RMS) 
& $1066 \pm 84$ & $5607 \pm 315$ & $4305 \pm 422$
\enddata 
\tablenotetext{a}{Widths are in the rest frame of NGC 7469.
\civ\ line widths are from \cite{Peterson04} and \cite{Collin06}.}
\label{Table:vwidths}
\end{deluxetable} 

\begin{deluxetable*}{lccccc} 
\tablewidth{0pt} 
\tablecaption{Virial Products} 
\tablehead{ 
\colhead{Emission} &
\colhead{} & 
\multicolumn{2}{c}{Cross Correlation (FR/RSS)} & 
\multicolumn{2}{c}{{\tt JAVELIN}}  \\
\colhead{Line} &
\colhead{Reference} &
\colhead{$\tau_{\rm cent}$} & 
\colhead{$M_{\rm vir}(\times 10^6\Msun$)} & 
\colhead{$\tau_{\tt JAVELIN}$} & 
\colhead{$M_{\rm vir}(\times 10^6\Msun$)}\\
\colhead{(1)} & 
\colhead{(2)} & 
\colhead{(3)} &
\colhead{(4)} &
\colhead{(5)} &
\colhead{(6)} 
}
\startdata
\Hbeta & 1 
&$10.8^{+3.4}_{-1.3}$ & $3.41^{+1.27}_{-0.79}$ & $9.8^{+1/2}_{-0.4}$ & $3.11^{+0.73}_{-0.63}$ \\
\heii\,$\lambda 4686$ & 1 
& $1.3^{+0.9}_{-0.7}$ & $1.30^{+0.93}_{-0.67}$ & $0.8^{0.7}_{-0.8}$ &  $0.86^{+0.76}_{-0.22}$ \\
\civ\,$\lambda1549$ & 1,2,3 
&$2.5^{+0.3}_{-0.2}$  & $3.35^{+0.50}_{-0.40}$ & $2.3^{+0.3}_{-0.3}$ & $3.13^{+0.49}_{-0.39}$ \\
\siiv\,$\lambda 1400$ & 2,3   
& $1.7^{+0.3}_{-0.3}$ & $4.05^{+0.95}_{-0.95}$ & $2.0^{+0.4}_{-0.5}$  & $4.77^{+1.20}_{-1.40}$    \\
\heii\,$\lambda1640$& 2,3   
& $0.6^{+0.3}_{-0.4}$ & $1.62^{+0.82}_{-1.09}$ & $0.8^{+0.2}_{-0.2}$  & $2.16^{+0.56}_{-0.56}$   \\
\Hbeta & 3,4 
& $4.5^{+0.7}_{-0.8}$ & $1.90^{+0.61}_{-0.63}$ & $\ldots$& $\ldots$ \\ 
\Halpha &3,4  
&$4.7^{+1.6}_{-1.3}$ &  $1.24^{+0.45}_{-0.37}$ & $\ldots$& $\ldots$ \\
Weighted Mean&  & $\ldots$ &$2.22 \pm 0.24$ &$\ldots$& $2.46\pm 0.26$  
\enddata 
\tablenotetext{1}{This work.}
\tablenotetext{2}{\cite{Wanders97}.}
\tablenotetext{3}{\cite{Peterson04}.}
\tablenotetext{4}{\cite{Collier98}.}
\label{Table:virial}
\end{deluxetable*}

\begin{figure}
\begin{center}
\epsscale{1.0}
\plotone{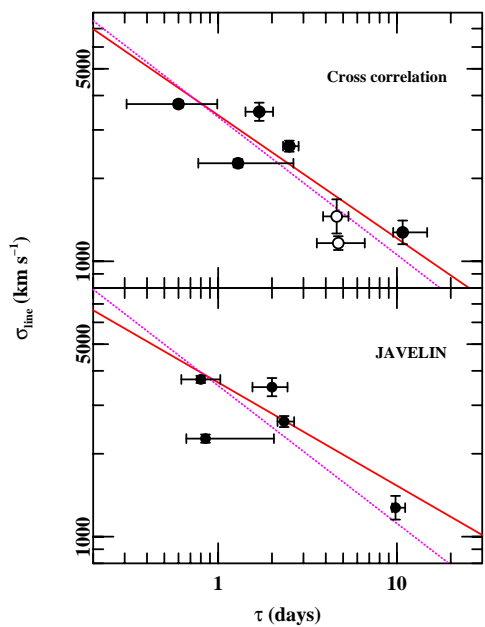}
\caption{The relationship between emission-line width and
lag. The top panel shows the relationship between line dispersion in the
RMS spectrum and cross-correlation centroid $\tau_{\rm cent}$. The
open circles are lower-confidence measurements of the \Hbeta\
and \Halpha\ lines from the AGN Watch program in 1996.
The red line is the fit to the relationship
$\log \sigma_{\rm line} = a + b \log\tau$, which 
has slope $b = -0.45 \pm 0.05$. 
The dotted magenta line is the best-fit with a forced
virial slope of $b = -0.5$.
The bottom panel is exactly the same as the top panel, 
except that the time lags $\tau$ are from {\tt JAVELIN}.
The {\tt JAVELIN} results for \Halpha\ and \Hbeta\ from the
AGN Watch program are ambiguous and therefore not included.
The best-fit slope for these data is $b = -0.38 \pm 0.06$.}
\label{fig:virial}
\end{center}
\end{figure}

A necessary condition for using reverberation results to estimate
black hole masses is that the virial products for the various emission
lines are the same. In Figure~\ref{fig:virial}, we plot line width
$\sigma_{\rm line}$ versus time delay for both $\tau_{\rm cent}$ and
$\tau_{\tt JAVELIN}$ from the detrended light curves
and find that the data are now generally quite 
consistent with the simple virial prediction.

All of our ignorance of the  geometry, kinematics, and inclination of the BLR
is subsumed in the scaling factor $f$ that is needed to 
convert $M_{\rm vir}$ into $M_{\rm BH}$. At the present time, it is
difficult to determine $f$ for an individual source, although
dynamical modeling of reverberation data is beginning to
show great promise in this regard \citep{Pancoast12, Pancoast14}.
In the absence of a determination of $f$ for a specific source such 
as NGC\,7469, we can determine an
ensemble average value $\langle f \rangle$ 
by employing a secondary method to estimate
AGN black hole masses. The commonly used method is to
assume that the relationship between central black hole mass and 
host-galaxy bulge velocity dispersion, the 
$M_{\rm BH}$--$\sigma_*$ relationship, is the same in active and quiescent
galaxies \citep{Onken04}. The most recent determination of 
the scale factor for reverberation-mapped AGNs is
$\langle f \rangle = 4.31 \pm 1.05$ \citep{Grier13b}. 
This estimate is consistent with
recent results by \cite{Woo10} and \cite{Park12}, who obtain estimates
of $\langle f \rangle$ = 5.2 and $\langle f \rangle$ = 5.1,
respectively, but it is about a factor of two larger than the
value of $\langle f \rangle$ computed by
\cite{Graham11}. \cite{Park12} attribute this factor of two difference
in $\langle f \rangle$ estimates to sample selection and to the regression method
used for the calculations. 

We estimate the mass of the black hole in NGC 7469
by using the weighted mean virial products in Table~\ref{Table:virial}
and by taking $f = 4.31$ \citep{Grier13b}. This yields masses of
$9.57\ (\pm 1.03) \times 10^6\,\Msun$ for the cross-correlation based results
and $10.6\ (\pm 1.12) \times 10^6\,\Msun$ for the {\tt JAVELIN} measurements.
The formal errors quoted here are the random components only, from the
uncertainties on the lag and line width. The systematic error, estimated from
the scatter around the $M_{\rm BH}$--$\sigma_*$ relationship, is probably
$\sim 0.43\,{\rm dex}$ \citep{Woo10}.

\section{DISCUSSION AND SUMMARY}

As we noted in \S\ref{section:intro}, the earlier reverberation
data on NGC\,7469 were not especially good because the signal-to-noise
of the UV data was suboptimal and the temporal
sampling of the optical data
was rather poor. Consequently, it was not too surprising that
NGC\,7469 was a significant outlier in the otherwise fairly tight
relationship between AGN luminosity and \Hbeta\ lag \citep{Bentz09}, 
the AGN BLR ``radius--luminosity relation.'' The new \Hbeta\ measurement
places NGC\,7469 within the scatter of this relationship.

We noted earlier that there are features in the emission-line light
curves that are seen clearly in the emission-line light curves, but not
in the continuum light curve. Specifically, these can best be seen in the 
detrended \heii\ light curve around HJD2455515 and
HJD2455530 and the detrended \Hbeta\ light curve around HJD2455540
in Figure~\ref{fig:dtlc}. We speculated that this might be a consequence
of the some sort of smoothing or reprocessing of the ionizing continuum
that occurs on timescales longer than the light-travel time between the
locations where the ionizing and optical continua are produced.  
As an experiment, we used the general methodology of {\tt JAVELIN} to
model the observed line and continuum light curves as 
differently lagged and smoothed versions of an unobserved underlying
UV continuum. These experiments were not successful. Either it is
possible  for short-time scale fluctuations to modify the line fluxes
without affecting the continuum or some of the short time scale structure
in the line light curves is due to an unappreciated systematic error
in their construction.

In summary, on
account of the low level of variability in this campaign, we were 
unable to meet our primary goal of recovering velocity--delay maps for
\Hbeta\ and \heii, as we did for other AGNs observed in the same
campaign \citep{Grier13a}, despite the intensive observational coverage.
We were, however, able to recover emission-line lags for these two
lines, but only after detrending the light curves. We also applied
detrending to the UV continuum and \civ\,$\lambda1549$ light curves 
from \cite{Wanders97} and thus resolved the discrepancy between
the \civ\ lags measured by cross-correlation and {\tt JAVELIN}
analyses of \citep{Zu11}. From these data, we able to derive a black hole mass
of $\sim 1\times 10^{7}\,\Msun$ for the central black hole, using
the most recent calibration of the reverberation mass scale
\citep{Grier13b}.

\acknowledgments BMP, CJG, GDR, and RWP are grateful for the support of the
National Science Foundation through grant AST-1008882 to The Ohio
State University. MCB gratefully acknowledges support from the NSF
through CAREER grant AST-1253702.
KDD acknowledges support by the NSF through
award AST-1302093 and from the Marie Curie Actions of the European 
Union's Seventh Framework Programme FP7/2007-2013/ under REA grant
agreement No. 300553. BJS, CBH, and JLV acknowledge support by NSF
Fellowships. CSK, AMM, and DMS acknowledge the support of NSF grants
AST-1004756 and AST-1009756. SK is supported at the Technion by the
Kitzman Fellowship and by a grant from the Israel-Niedersachsen
collaboration program. SR is supported at Technion by the Zeff
Fellowship. SGS acknowledges the support to CrAO in the frame of the
`CosmoMicroPhysics' Target Scientific Research Complex Programme of
the National Academy of Sciences of Ukraine (2007-2012). VTD
acknowledges the support of the Russian Foundation of Research (RFBR,
project no. 12-02-01237-a). The CrAO CCD cameras were purchased
through the US Civilian Research and Development for Independent
States of the Former Soviet Union (CRDF) awards UP1-2116 and
UP1-2549-CR-03. This research has been partly supported by the
Grant-in-Aids of Scientific Research (17104002, 20041003, 21018003,
21018005, 22253002, and 22540247) of the Ministry of Education,
Science, Culture and Sports of Japan. This research has made use of
the NASA/IPAC Extragalactic Database (NED), which is operated by the
Jet Propulsion Laboratory, California Institute of Technology, under
contract with the National Aeronautics and Space Administration.

\end{document}